\documentclass[%
 reprint,
superscriptaddress,
nofootinbib,
amsmath,amssymb,
 aps,
]{revtex4-2}

\usepackage{lipsum}
\usepackage{xcolor}
\usepackage{mathrsfs,amsmath}
\usepackage{gensymb}
\usepackage{ulem}
\usepackage{xcolor}
\usepackage{graphicx}
\usepackage{dcolumn}
\usepackage{bm}
\usepackage[hidelinks]{hyperref}
\usepackage{miller}
\newcommand{\cmmnt}[1]{\ignorespaces}

\makeatletter
\renewcommand*{\@fnsymbol}[1]{\ensuremath{\ifcase#1\or \dagger \else\@ctrerr\fi}}
\makeatother

\begin{document}

\preprint{APS/123-QED}

\title{Probing negative differential resistance in silicon \\ with a P-I-N diode-integrated T center ensemble}

\author{Aaron M. Day}
    \affiliation{John A. Paulson School of Engineering and Applied Sciences, Harvard University, Cambridge, Massachusetts 02138, USA}
    \affiliation{Department of Physics, Harvard University, Cambridge, Massachusetts 02138, USA}
    \affiliation{Harvard Quantum Initiative, Harvard University, Cambridge, Massachusetts 02138, USA}
\author{Chaoshen Zhang}
    \affiliation{John A. Paulson School of Engineering and Applied Sciences, Harvard University, Cambridge, Massachusetts 02138, USA}
\author{Chang Jin}
    \affiliation{John A. Paulson School of Engineering and Applied Sciences, Harvard University, Cambridge, Massachusetts 02138, USA}
\author{Hanbin Song}  
    \affiliation{Department of Materials Science and Engineering, University of California, Berkeley, Berkeley, CA 94720, USA}
    \affiliation{Materials Sciences Division, Lawrence Berkeley National Laboratory, Berkeley, CA 94720, United States of America}
\author{Madison Sutula}
    \affiliation{Department of Physics, Harvard University, Cambridge, Massachusetts 02138, USA}
    \affiliation{IonQ Inc., 4505 Campus Dr., College Park, 20740, MD, USA}
\author{Denis D. Sukachev}  
    \affiliation{IonQ Inc., 4505 Campus Dr., College Park, 20740, MD, USA}
\author{Alp Sipahigil}
    \affiliation{Materials Sciences Division, Lawrence Berkeley National Laboratory, Berkeley, CA 94720, United States of America}
    \affiliation{Department of Electrical Engineering and Computer Sciences, University of California, Berkeley, CA 94720, United States of America}
    \affiliation{Department of Physics, University of California, Berkeley, CA 94720, United States of America}
\author{Mihir K. Bhaskar}  
    \affiliation{IonQ Inc., 4505 Campus Dr., College Park, 20740, MD, USA}
\author{Evelyn L. Hu}
    \thanks{Corresponding Authors: \\ aday@g.harvard.edu, \\ ehu@seas.harvard.edu}
    \affiliation{John A. Paulson School of Engineering and Applied Sciences, Harvard University, Cambridge, Massachusetts 02138, USA}
    \affiliation{Harvard Quantum Initiative, Harvard University, Cambridge, Massachusetts 02138, USA}

\date{\today}
\begin{abstract}
    Solid-state defect quantum systems are exquisite probes of their local charge environment. Nonlinear dynamical electric fields in solids are challenging to characterize directly, conventionally limited to coarse macroscopic methods which fail to capture subtle effects in the material. Here, through transient optical spectroscopy on an embedded T center ensemble, we realize the in-situ observation of a silicon PIN-diode phase transition to a regime of self-sustained carrier oscillatory dynamics characteristic of negative differential resistance. Manifest in both the ensemble electroluminescence and photoluminescence, we find a temperature and field-dependent phase space for persistent undamped amplitude oscillations indicative of a collective ensemble response to the field dynamics. These findings shed new light on the cryogenic behavior of silicon, provide fundamental insight into the physics of the T center for improved quantum device performance, and open a promising new direction for defect-based local quantum sensing in semiconductor devices.
\end{abstract}
\maketitle

\section{Main}
Solid-state defects show promise for a multiplicity of quantum information applications, ranging from high-sensitivity in-situ probes to building blocks of quantum networks. Recent progress in defect integration with nano- photonic, phononic, and electronic devices demonstrate the maturity of defect-device enhancement for improved readout, control, and sensing \cite{riedel2025,ding2024,dietz2023,anderson2019}. In particular with electrical devices, recent attention has focused on imparting tunable control over defect optical and spin properties for improved performance in quantum networks \cite{anderson2019, lukin2020stark, wirtitsch2024, lohrmann2015, dobinson2025}. However, defects conversely provide a wealth of detailed information regarding their surrounding environment--including band edge proximity, charge state occupancy, spin bath, and local charge and magnetic noise \cite{day2024,widmann2019,dolde2011}. The silicon T center is a particularly interesting test case for electronic integration, as its optical transition is believed to be due to a bound exciton, with one state close to the conduction band edge \cite{dhaliah2022}. Additionally, the T center is a leading quantum information candidate owing to its optically-accessible ground state spin with modest coherence time, nuclear spin coupling, telecommuncation-band optical transition, and commercial substrate device integration \cite{higginbottom2022,komza2024,song2025,dobinson2025}.

Although defect electronic integration is an active field \cite{anderson2019,day2024,dobinson2025,steidl2025,zeledon2025}, defect-mediated investigation of nonlinear dynamics is an underexplored opportunity. Negative differential resistance (NDR) is one such well-known solid-state electrical nonlinear phenomenon. The first observations of self-sustained nonlinear oscillatory behavior linked to NDR were in biased gallium arsenide, known as the Gunn Effect \cite{gunn1964}. The physics underlying Gunn NDR is a field-dependent carrier deceleration resulting from promotion to a neighboring conduction band level with different effective mass \cite{kroemer1964}. This band-mediated example of NDR is a subset of a broader class of transferred-electron phenomena in which carrier nonlinearities can occur from many varied underlying physics. NDR typically exhibits an inversion in the current-voltage (IV) relation, revealing a region where  $\frac{dI}{dV} < 0$, with $\frac{dI}{dV} > 0$ regions surrounding it at lower and higher field strengths. This produces an instability within the NDR bias region with an associated oscillatory nonlinearity. A generalized framework understanding NDR as a non-equilibrium phase transition later emerged \cite{landsberg1976}, where semiconductor nonlinearities have now been studied for many decades \cite{scholl2001,scholl2012}. Significant effort has since been dedicated to this field, though restricted to conventional experimental methods of IV-measurements \cite{dudeck1977,kassing1975}, electron-beam-induced-voltage scanning-electron microscopy \cite{baumann1986}, and potential-probe surface conductivity \cite{symanczyk1991}, which are limited in spatial, temporal, and absolute field sensitivity.  

In contrast, here we establish the ability of excitonic color centers in silicon to serve as atom-scale probes to the local nonlinear electrical environment--directly mapping the transient electrical nonlinearities onto telecom optical photons thereby enabling optical observation of the semiconductor phase transition. We couple an isolated $50~\mu$m ensemble of silicon T centers to a lateral P-I-N diode fabricated via lithographically-defined ion-doping and probe the defect response to applied electric fields. We reveal a field- and temperature-dependent regime for self-sustained stable oscillations in current and T-center luminescence induced by forward biasing. These results suggest an electric field-induced phase transition characteristic of NDR-mediated by impurity trapping/release, and impact ionization \cite{hupper1993,scholl2001,scholl2012}, resulting in oscillatory diode current which couples to the T center. These findings expand the understanding of the fundamental behavior of the silicon T center, and motivate the further exploration of defect-enabled quantum sensing and metrology in semiconductors. 

\subsection{P-I-N diode-integrated T Center}
\begin{figure}[ht!]
\includegraphics[scale = 1]{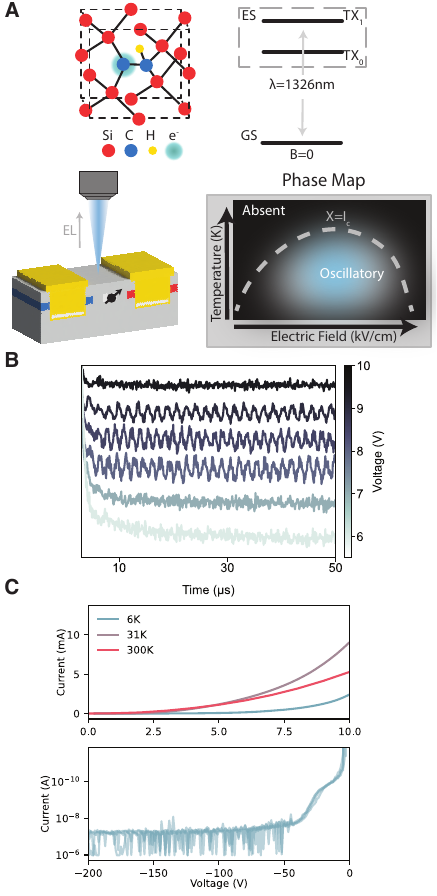}
\caption{\textbf{DC-bias induced electrical phase transition coupled to emission of telecom photons from silicon artificial atoms} \textbf{(a.)} The T center is a carbon-related defect in silicon which emits photons in the telecommunication-band and possesses an addressable ground-state spin. An ensemble of T centers is incorporated in a buried lateral P-I-N diode fabricated in silicon via lithography-defined ion doping. Arrow indicates T center electroluminescence (EL) emission upon applied forward bias. MHz-frequency carrier oscillations are induced via application of a forward DC-bias which modulate the optical emission of an embedded T center ensemble. The oscillatory electrical dynamics are characteristic of a temperature- and field-dependent phase transition from the diode-ensemble interaction, observed when the impact ionization control parameter X crosses a critical point $I_c$ \cite{scholl2012}. \textbf{(b.)} Field-dependence of oscillatory T center ensemble photoluminescence (PL) at 6~Kelvin. \textbf{(c.)} IV curves of P-I-N diode with integrated T center ensemble. Reverse-bias reveals burst noise characteristic of defect trapping \cite{hsu1970}, with multiple traces provided illustrating the random fluctuations.}
\label{Fig1}
\end{figure}

The T center is a carbon-related O-Band-emitting defect in silicon with an optically-accessible ground-state spin. The defect atomic complex is comprised of two bonded carbon atoms, one of which has a paired hydrogen atom, substituting a silicon site (Fig. \ref{Fig1}a). The T center emission process is mediated by recombination of a bound exciton localized at the defect complex \cite{xiong2024_1,xiong2024_2}, resulting in disassociation at elevated temperatures ($\textgreater$40~K) and a weak optical transition with long excited state lifetime ($\sim1\mu$s). The T center was discovered in the late 20th century among a class of semiconductor impurities \cite{minaev1981,safonov1996,hayama2004}, but has recently gained new interest for its potential as a quantum memory \cite{bergeron2020, higginbottom2022}. 

We localize a small ensemble of T centers to the center of a lithographically-defined ion-doped P-I-N diode at a buried plane in a high-resistivity intrinsic-type silicon wafer (Fig. \ref{Fig1}a). The device studied in this work possesses a $565~\mu$m-wide \textit{I-}region, though the width per device is varied across the wafer. The processing (and step order) for emitter formation and diode fabrication are carefully chosen to ensure successful creation of both the electrical device and the integrated defect complex (Methods). Using this device, the dynamic-response of T center photoluminescence and electroluminescence are explored.  Surprisingly,  we observe that the application of a DC-bias to our P-I-N junction results in the induction of radio-frequency oscillations of propagating carriers which directly modulate the optical emission from the T center (Fig. \ref{Fig1}b), attributed to a phase transition in the electrical dynamics of the device.

Forward-biasing a diode efficiently drives electron and hole current across the device from their respective N- and P- contacts which can be captured at a defect trap site and induce electroluminescence (EL) \cite{piper1955}. The conventional forward-bias characteristic results from voltage-controlled lowering of the barrier between N- and P- regions, allowing the cross-junction transport of electrons in the conduction band and holes in valence band. At low temperatures such as 10~K, a number of nuances arise. The very low ionization of the dopant levels restricts the electron and hole concentrations \cite{jonscher1961}. The IV-relation of the device in forward bias (Fig. \ref{Fig1}c) thus shows diminished total current, and a higher voltage turn-on as the temperature varies from room-temperature to 6~K cryostat base. As such, a portion of the current may emanate from injection of non-thermalized electrons and holes across the metal-semiconductor interface. Further, the IV-curve reveals burst-noise at 6~K under high reverse-bias field. Burst noise in semiconductors has been attributed to defects in the material, and has been correlated with random trapping and release of charge from those defects \cite{hsu1970}. The burst-noise was not present in similar devices possessing a G center ensemble \cite{day2024}. This signature of trap-state interaction is explored at length in Section C. 

\subsection{Photoluminescence and Electroluminescence}
\begin{figure}[ht!]
\includegraphics[scale = 1]{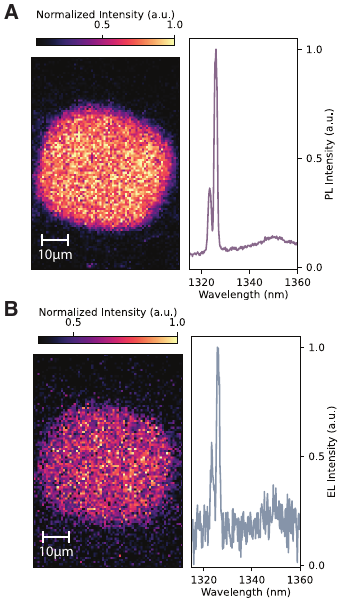}
\caption{\textbf{T center DC-bias optical response} Intensity confocal scans and zero phonon line spectra of lithographically-defined T center ensemble confined to the I-region of a P-I-N diode measured via  \textbf{(a.)} photoluminescence ($1~$mW, $532~$nm excitation), and \textbf{(b.)} electroluminescence ($9~$V, $\sim2~$mA). Relative PL vs EL intensity curves given in supplement \cite{SI}.}
\label{Fig2}
\end{figure}

We first characterize the steady-state forward-biased T center-diode interaction at cryostat base temperature ($\sim 6~K$) in Fig. \ref{Fig2}. The un-biased optical properties of the diode-integrated ensemble are identified using photoluminescence (PL) confocal scans (Fig. \ref{Fig2}a). The spatial localization of the $50~\mu$m ensemble--as defined by the lithographic mask for isolated hydrogen implantation--is evident under 532~nm laser excitation, with no emitters present outside the defined region. T center TX$_0$ and TX$_1$ zero phonon line (ZPL) optical transitions are uniformly observed in the spectra of the confocal scan. 

A positive bias is then applied to the diode to pass electron and hole currents across the ensemble, while a scanning mirror generates a confocal map of electroluminescence (EL). With the laser turned off, T center emission is apparent in both the spatial confocal scan and the spectra (Fig. \ref{Fig2}b), matching that of the PL scan (where relative PL/EL saturation is reported in SI \cite{SI}). The ensemble EL is driven below saturation due to current-induced heating typical of inefficient capture across a large diode area. However, spatial carrier confinement and waveguide extraction enables an efficient electrically-gated T center as recently demonstrated \cite{dobinson2025}. 

\subsection{Transient Oscillations}

\subsubsection{DC-Biased PL Dynamics}
\begin{figure}[h!]
\includegraphics[scale = 1]{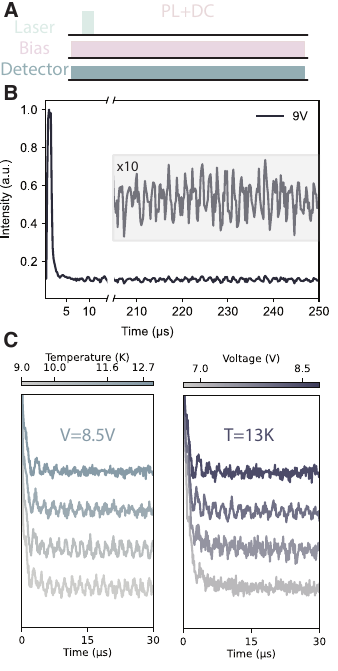}
\caption{\textbf{Transient oscillatory dynamics in DC-biased photoluminescence} \textbf{(a.)} Pulse sequence for DC-biased pulsed photoluminescence.  \textbf{(b.)} Pulsed PL+DC oscillations. Within the stable region (9V, 6~Kelvin), oscillations persist without damping. \textbf{(c.)} Mapping the oscillation instability as a function of temperature (left) and field (right) while holding constant the alternative control parameter.}
\label{Fig3}
\end{figure}

In Fig. \ref{Fig3}, we study the dynamics of T center pulsed PL under a constant DC forward bias (Fig. \ref{Fig3}a). The ensemble emission possesses an undamped oscillatory component which lasts beyond the measurement window (500$\times$ the optical lifetime) (Fig. \ref{Fig3}b). While at the base-temperature of 6~K, we sweep the magnitude of DC-bias applied to the junction and find a narrow voltage range which supports undamped oscillations, whereas above and below this voltage condition the emitter decay ceases to possess oscillations. We then probe the instability of the observed phenomenon in Fig. \ref{Fig3}c, sweeping the cryostat temperature while at a voltage of stable oscillations (left), and sweeping the voltage while at a temperature of damped oscillations (right). Similar damping dynamics are exhibited in the oscillations of both regimes, suggesting temperature and field may be distinct dependent variables which control the effect. Under this protocol, the optically detected signal exhibit higher SNR than via electrical detection. We emphasize the peculiarity of observing an AC-component in the defect emission under static electrical DC-bias in a conventional silicon P-I-N diode--and the narrow energy band which supports it.

\subsubsection{EL Dynamics}
Here we report characterization of T-center diode behavior under fast switching. We utilize a forward-biased variable-amplitude square-wave from an arbitrary waveform generator. During pulse sequence optimization we observe a slow charge accumulation in the junction which is effectively mitigated with reverse-bias, analogous to erasure of non-volatile memory \cite{pavan1997} (see supplemental information (SI) \cite{SI}). Therefore we follow the forward-bias pulse with a constant reverse bias for the duration of the pulse period. We construct a measurement circuit to directly probe the transient current in the diode (see SI \cite{SI}), to compare the electrical dynamics to those observed in the emitter. In Fig \ref{Fig4}a-b, we analyze the transient response of the (a) T center against that of the (b) diode current under varied magnitude of applied forward-bias voltage--where the fast fourier transform (FFT) is generated for each bias point (right). We establish both that the current in the diode itself is stably oscillating under these conditions, and that the T center optical emission directly mirrors these carrier dynamics--with an increased relative oscillation amplitude (see SI for further \cite{SI}). Observation of diode current oscillations in this pulsed EL protocol is unique, as only optical oscillations could only be detected in the static DC-bias protocol of Fig. 3. Furthermore, the oscillation frequencies find near-perfect agreement between the diode current and T center emission.

\subsubsection{Oscillation Phase Space}
\begin{figure*}[ht!]
\includegraphics[scale = 1]{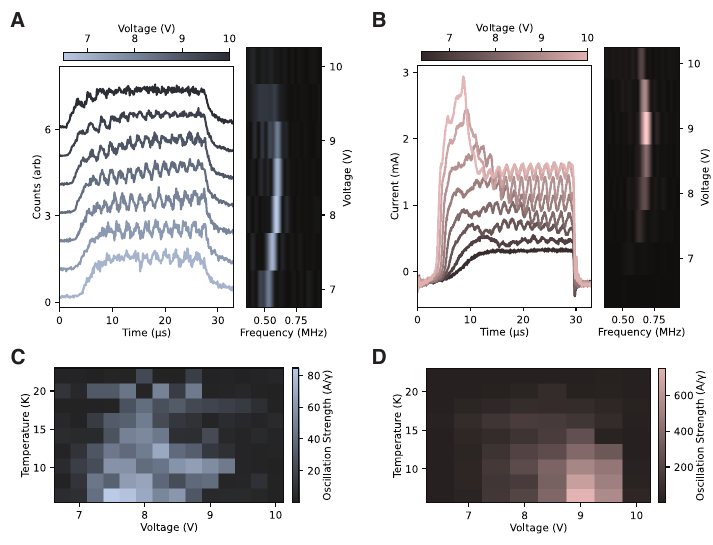}
\caption{\textbf{Phase space of oscillation stability} \textbf{(a.)} Voltage dependence of transient \textit{EL} oscillations observed in T center optical response (left) with FFT per voltage (right) \textbf{(b.)} Voltage dependence on transient \textit{diode current} oscillations (left) with FFT per voltage (right). The T center emission is vertically offset for clarity, whereas the vertical rise in diode current is directly due to the increased applied voltage. \textbf{(c.)}-\textbf{(d.)} Phase map of oscillation strength as a function of voltage and temperature, revealing stable and absent phases of oscillations in the T center (c), and diode current (d). The oscillation strength $A/\gamma$ (where $A$ and $\gamma$ are the fourier transform amplitude and linewidth, respectively) at each voltage and temperature condition is measured by taking the fourier transform of the oscillation dynamics. A phase transition is observed in the traversal across the x- and y-axes of the phase diagrams, crossing a critical point of the impact ionization control parameter.}
\label{Fig4}
\end{figure*}

Based on the findings of Fig. \ref{Fig3}c and  Fig. \ref{Fig4}a-b, we postulate that the creation of stable domains of electric-field oscillation may be described with a phase-diagram. We thus repeat the experiment of Fig. \ref{Fig4}a-b at increments of voltage and temperature, extracting the oscillation strength--given by the FFT amplitude and linewidth--per condition. Sweeping the forward-bias voltage from 6.75-10~V and temperature from 6-22~K, we map a phase-diagram with a transition from stable-damped-absent electric field oscillations of the T center (Fig. \ref{Fig4}c) and the diode (Fig. \ref{Fig4}d). Indeed, we again find strong agreement between the response of the emitter and the diode--with strongest stable oscillations occurring between 8-9~V and 6-13~K (see SI for complete FFT dataset \cite{SI}). 

We emphasize that the observed stable oscillations cannot be explained simply by parasitic electrical inductance and capacitance, as ringing of that kind is not indicated by undamped oscillatory current under a static DC-bias, the T center lifetime measured under fast electrical switching is identical to that of pulsed-PL, and the circuit decay is order-magnitude faster than the EL lifetime (SI \cite{SI}). Additionally, we confirm that the optical source of the oscillations is uniquely that of the T center--rather than some other background defect--through careful spectral filtering and analysis in the region away from the T-center ensemble \cite{SI}. Rather, a property of the diode itself--probed directly with the T center--instantiates the observed phenomena. Indeed, there exists a basis in the literature for such anomalous behavior in semiconductors--that of impurity-mediated impact ionization \cite{hupper1993, scholl2001, scholl2012}.

\subsection{Discussion}
\subsubsection{Origin of EL}
Electroluminescence in a bulk-semiconductor diode-structure is enabled by a number of mechanisms. Free carriers may be generated through direct impurity ionization under electric field, impact-ionization of impurities through accelerated injected carriers \cite{smith1975,weman1986}, and injection of electrons and holes from the respective N- and P-doped regions \cite{piper1955}. Depending on the energy level and excitation pathways, defects in the bandgap can capture these free carriers for promotion to and de-excitation from an excited state. For excitonic defects--the class to which the T center belongs--electroluminescence may assist us in understanding the formation and capture of excitons. EL could involve (1) injection of electron- and hole-current from the respective N- and P-doped regions recombine to form excitons in the \textit{I-} region \cite{palm1996}, and/or (2) the impact-ionization of local impurities under field-mediated acceleration of carriers may form free-excitons \cite{eitan1982,maes1990}. Once formed, the free-excitons can then be captured and recombine at the defect site driving EL, similar to the proposed method of optical above-bandgap T center excitation \cite{xiong2024_1,xiong2024_2}.

\subsubsection{Origin of Oscillations}
To better understand induced carrier nonlinearities under DC-biasing, it is helpful to consider the expression for current density in a semiconductor--expressed with Equation 1:
\begin{equation}
\vec{J}=q(\mu_n {\bf n} + \mu_p {\bf p}) \vec{E}
\end{equation}
where q is the electron charge, {\bf n, p} are the densities of the free electrons and holes, $\mu_n$ and $\mu_p$ are the corresponding mobilities, the average carrier velocity is $\vec{v_{n,p}} = \mu_{n,p} \vec{E}$, and $\vec{E}$ is the applied electric field. {\bf n, p}, $\mu_n$ and $\mu_p$ are all temperature-dependent quantities. 

Numerous mechanisms permit field-dependent oscillatory carrier density. The electron or hole mobilities, $\mu_n$ and $\mu_p$ can be modified under increased field--as with the canonical GaAs Gunn oscillations where increased field scatters electrons into different effective mass valleys. However more generally NDR describes phenomena which simply induce inhomogeneities in the current density and/or field distribution, resulting in a decreased current density at increased electric field \cite{scholl2012}.  Such inhomogeneity may be expressed by Equation 2: 
\begin{equation}
    \vec \nabla \cdot \vec J \ne 0 
\end{equation}
where the divergence of the current density is non-zero due to nonlinearities in carrier-density and/or mobility.  

Impurity states within the bandgap are well understood to give rise to such inhomogeneities; these influence the lifetimes of free electrons and holes, and hence modify the concentration and dynamics of those free carriers. Indeed, a number of early studies on electrical nonlinearities in semiconductors identified mid-gap defect trap states from Au-doping as the origin of NDR in Si PIN-diodes at room temperature, and similar phenomena has been observed in Germanium diodes at liquid helium temperature \cite{lampert1961,weber1970, deuling1970,kassing1975,dudeck1977,jager1986,baumann1986,weman1989,symanczyk1991, Mayer1987}. The oscillatory behavior of NDR results from the continued cycling of electrons back and forth between occupied valleys (for the Gunn effect) or between trapped and un-trapped carriers (in our case).

\subsubsection{Phase Transition Picture}
A key to the observation of oscillatory behavior is a non-linear coupling between the products of carrier generation and recombination \cite{scholl1989theoretical}. One such process is the impact ionization of impurity levels inducing an avalanche of carriers in the conduction or valence band, competing with a reduction in carrier density via the capture and recombination involving excitons. The phase transition is tuned via a temperature- and field-induced increase in carrier concentration (generation, {\bf g}) through impact ionization, and a decrease in carrier concentration through recombination ({\bf r}) that produce charge-state alteration of defect states and changes in trapping lifetimes. The impact ionization rate is further modified by the energy relaxation of non-equilibrium carriers. Utilizing impact ionization as the autocatalytic mechanism, Landesberg and Pimpale \cite{landsberg1976} have proposed a model with a temperature and field-dependent control parameter $X_c = C_i(T,E)-l$, where $C_i(T,E)$ is the impact ionization coefficient, and $l$ is the inverse of a characteristic recombination lifetime. The transition (oscillation onset) occurs depending on the value of the control parameter, with the electron concentration as the order parameter. Weman, Henry and Monemar observed impact ionization-induced spontaneous oscillations in silicon \cite{weman1986}, and postulated their frequency origin as a rate competition between impact ionization break-up and subsequent re-formation of excitons. 

The condition on temperature and voltage that bound the phase transitions we observe in Fig.~\ref{Fig4}(c) and (d) suggest important variables that influence charge injection and capture into the region of the T centers. Evidenced by the spatial and spectral signature of our observed oscillations uniquely emanating from the T center ensemble itself (see SI for further \cite{SI}), we postulate that the T centers themselves--introduced from high-density ion implantation in a (ultrapure) PIN-junction--could be contributing to the generation-recombination (g-r) processes. We thus believe the framework of recombination-induced phase transitions can provide a starting point for the further modeling of the behavior of our T center-integrated lateral P-I-N diodes. 

\section{Conclusion and Outlook}
We have demonstrated the direct probing of nonlinear electrical dynamics inside a silicon P-I-N diode by analyzing an embedded T center ensemble's transient spectroscopy. Measuring T center carrier capture under pulsed electroluminescence and DC-biased photoluminescence, we map a nonequilibrium phase transition in the electrical nonlinearities which manifest in the emitter optical emission. We find strong agreement between the direct measure of transient electrical current and EL oscillations, whereas in the DC-biased pulsed PL regime we observe phenomena which cannot be detected in the current--suggesting a defect-enhanced sensitivity to a weaker effect in the semiconductor. 

Many exciting questions remain from these findings. The extent to which the T centers merely sense oscillatory nonlinearities, as opposed to cause them, remains an open question. Indeed, the predicted near-band edge position of the T center level \cite{dhaliah2022} may play a contributory role. Varying the T center ensemble density and material properties (resistivity, purity) could inform whether the trap states introduced by the T centers directly cause the observed electrical nonlinearities. Additionally, further studies are needed to understand the distinction between the two regimes we study--DC-biased PL vs switched-EL--as a unique second-order phase transition is predicted for low-temperature stimulated-exciton creation which may more fully capture our DC-biased PL protocol \cite{landsberg1976}. Finally, the robust time-synchronized and persistent oscillatory emission of an ensemble is striking, as one may instead expect dephasing from random fluctuations in the capture dynamics of the constituent emitters across the ensemble. Our findings reveal striking observations regarding the cryogenic nonlinear dynamics of silicon PIN diodes, and opens new possibilities for control and read-out in the sensing of nonlinear fields. 

\section{Methods}
\subsection{Emitter Formation and Diode Fabrication}
T centers are formed utilizing a series of carbon-ion implantation, hyrdrogen-ion implantation, and rapid thermal anneals--the exact details of which are given below. In our investigation of emitter synthesis, we found the T center formation to be sensitive to processing conditions, and thus great care is taken in particular in the thermal annealing steps. Additionally, other work has utilized a water-boiling step for T center formation, whereas we do not need this step. Due to the T center processing sensitivity, the steps for diode fabrication (also given in detail below) are assessed and interspersed accordingly to co-form the device and emitter without degradation of each. A similar device fabrication protocol we used in a prior work \cite{day2024} is used here, modified to accommodate T center formation rather then G center. The full emitter formation and diode fabrication steps are given in order below. 

We begin with a 6-inch wafer of high-purity high-resistivity intrinsic-type silicon (Topsil, \textgreater10k$\Omega$-cm). First, the wafer is uniformly blanket-implanted with 7$\times10^{12}$/cm$^2$ $^{12}$C ions at an energy of 35~keV. Next, optical lithography is used to write an array of $500\times500~\mu$m aperture openings in a photoresist mask, and 1$\times10^{14}$/cm$^2$ $^{11}$B ions are implanted at an energy of 29~keV to define localized P-doped regions. The mask is removed, a new mask is added, an array of $500\times500~\mu$m apertures are exposed--aligned-to and offset-from the former p-dopant islands--and 1$\times10^{14}$/cm$^2$ $^{31}$P ions at an energy of 80~keV are implanted through the openings to constitute N-doped regions. The spacing between the P- and N-doped regions is defined and controlled to vary the width of the I-region of a formed lateral P-I-N diode. The mask is removed, and the wafer is rapid thermal annealed (RTA) at $\sim900^{\circ}$C for 20 seconds in an N$_2$ environment to all in one step heal the silicon lattice from ion-implantation damage and incorporate the boron and phosphorus ions substitutionally doping the lattice. Next, a grid of 250$\times$250$~\mu$m apertures are written in a new lithographic mask, with each square aligned to the center of the respective P- and N- doped regions, and a standard silicon-based reactive ion dry etch is performed using SF$_6$ and C$_4$F$_8$ chemistry, thereby etching the silicon down to expose the center of the doped regions. The etch mask is removed, and the wafer is now diced into  8$\times$8$~$mm samples--both in anticipation of the ultimate measurement apparatus geometry, and to improve calibration of subsequent thermal annealing. A new lithographic mask is created, with 50$\times$50$~\mu$m apertures defined at the exact middle (in x- and y-axes) of the formed lateral P-I-N junction, and 7$\times10^{12}$/cm$^2$ $^{1}$H ions are implanted at an energy of 8~keV. The mask is removed, and the sample is RTA at $\sim400^{\circ}$C for 180~s in an N$_2$ environment, thereby creating a small, dense ensemble of T centers localized to the center of a P-I-N diode. Finally, a grid of 300$\times$300$~\mu$m apertures, aligned to the etch-trenches, is lithographically defined in photoresist on lift-off (S1813 on LOR3A), and a layer of Ti-Au metal is electron-beam evaporated to form the diode metal contact. Filling an etched trench with metal ensures an ohmic metal-dopant-semiconductor interface for improved diode performance. The contacts are lifted-off using remover PG at $80^{\circ}C$. At this stage, both the device and emitter fabrication are complete. IV-characteristics of devices are pre-characterized at room temperature using a probe station, then mounted and wire-bonded to the cryostat electrical feed-through for cryogenic measurement. 

All carbon, boron, phosphorus, and hydrogen ion implantation is performed by Innovion Corporation (Coherent). All implantations are at a 7$^{\circ}$ tilt and with energies targeting peak concentration 110~nm below the sample surface (using Stopping Range of Ions in Matter (SRIM) calculations). Boron and phosphorus dopant densities are selected to achieve a peak acceptor/donor concentration of 1$\times 10^{19}$/cm$^3$, previously determined to yield optimal diode performance \cite{day2024}. Boron and phosphorus doping, and hydrogenation for T center formation, are localized to specific desired regions to define the diode geometry and ensemble location using lithographic masks. These masks are created with the positive photoresist S1813, written using approximately 250~mJ optical lithograhpy and 375~nm light on a Heidelberg Maskless Aligner (MLA 150). Before writing the resist is baked at $115^{\circ}C$ for 3 minutes, and after writing is developed for 70 seconds in TMAH-based CD-26. Lithographic masks are removed using Piranha and/or subsequent acetone-IPA rinses. 

\subsection{Experimental Setup}
Experiments are performed on a custom home-built free-space confocal microscope with a Mitutoyo 100x 0.5 NA Near-IR excitation objective, O-band collection optics, and above-bandgap excitation paths (see SI Section 2.0.5). Photons are collected and fiber routed to either a spectrometer (Acton Spectra Pro 2750 spectrograph with a Princeton Instruments OMA:V indium-gallium-arsenide nitrogen-cooled photodiode array detector) or O-band optimized superconducting nanowire single photon detectors (SNSPDs, PhotonSpot) housed on the 1~K stage of a dilution refrigerator. An acousto-optic modulator (AOM) is used to perform pulsed PL measurements of Fig. 2c and Fig. 3. We excite the sample with a laser power of 500$\mu$W and pulse-width of approx. 100~ns (noting that the AOM has a turn-off time of $\sim$50~ns).

The sample is mounted-on and wirebonded-to an electrically-wired cold-finger with a 16-pin removable electrical insert (see SI \cite{SI}) in a continuous-flow Janis ST-500 cryostat with integrated electrical feed-throughs for temperature sensing/control, and diode electrical biasing. Liquid helium is flowed from a dewar, circulated in the cryostat to reach a base temperature of roughly 6~Kelvin, and recovered in a university network. Temperature sweeps are performed via careful balance of helium flow rate and heating from a LakeShore 321 AutoTuning temperature controller (details in SI \cite{SI}). Steady state electrical-biasing of the diode is performed with a ±210~V Keithley 2400 sourcemeter. Transient electrical switching of the diode is performed with a Rigol IX1 DG 1022 arbitrary waveform generator. 

\section{Author contributions}
\noindent \textit{Methodology}: A.M.D., C.Z., E.L.H; \textit{Emitter Synthesis}: A.M.D. H.S., M.S.; \textit{Fabrication}: A.M.D.; \textit{Measurement}: A.M.D., C.Z., C.J.; \textit{Analysis}: A.M.D., C.Z., E.L.H.; \textit{Advising}: D.D.S., A.S., M.K.B., E.L.H; \textit{Manuscript Preparation}: All authors.

\section{Acknowledgments}
\noindent The authors thank Jonathan Dietz and Jim MacArthur for helpful conversations on optics and electronics, Yan-Qi Huan for assistance with dilution fridge installation, and Michael Haas for assistance with SNSPD and dilution fridge installation. 

\noindent This work was supported by AWS Center for Quantum Networking and the Harvard Quantum Initiative. Portions of this work were performed at the Harvard University Center for Nanoscale Systems (CNS); a member of the National Nanotechnology Coordinated Infrastructure Network (NNCI), which is supported by the National Science Foundation under NSF award no. ECCS-2025158. M.S. acknowledges funding from a NASA Space Technology Graduate Research Fellowship. H. S. and A. S. acknowledge funding from the U.S. Department of Energy, Office of Science, Basic Energy Sciences in Quantum Information Science under Award Number DE-SC0022289 for materials processing.

\section{Data Availability}
\noindent The data that support the findings of the work are available from the corresponding author upon reasonable request.

\bibliography{Ref.bib}
\end{document}


\title{Supplementary Information for ``Probing negative differential resistance in silicon with a P-I-N diode-integrated T center ensemble"}

\date{\today}
\maketitle

\tableofcontents
\clearpage

\section{Complete Frequency Map of Electroluminescence Phase Diagram}
Here we present the extended data used to generate the phase map of Main Text Fig. 4a. The pulsed EL protocol (SI Section IV) is utilized. At each temperature point, an applied voltage is swept and T center electroluminescence oscillations are measured per condition. The Fourier transform (FFT) is calculated from each time trace, and the intensity of oscillations is plotted as a heatmap per voltage. Signal padding is applied to the time trace to improve the FFT resolution. The complete dataset is given in Supplementary Fig. \ref{SIFig1}.

\begin{figure*}[ht!]
\includegraphics[scale = 0.9]{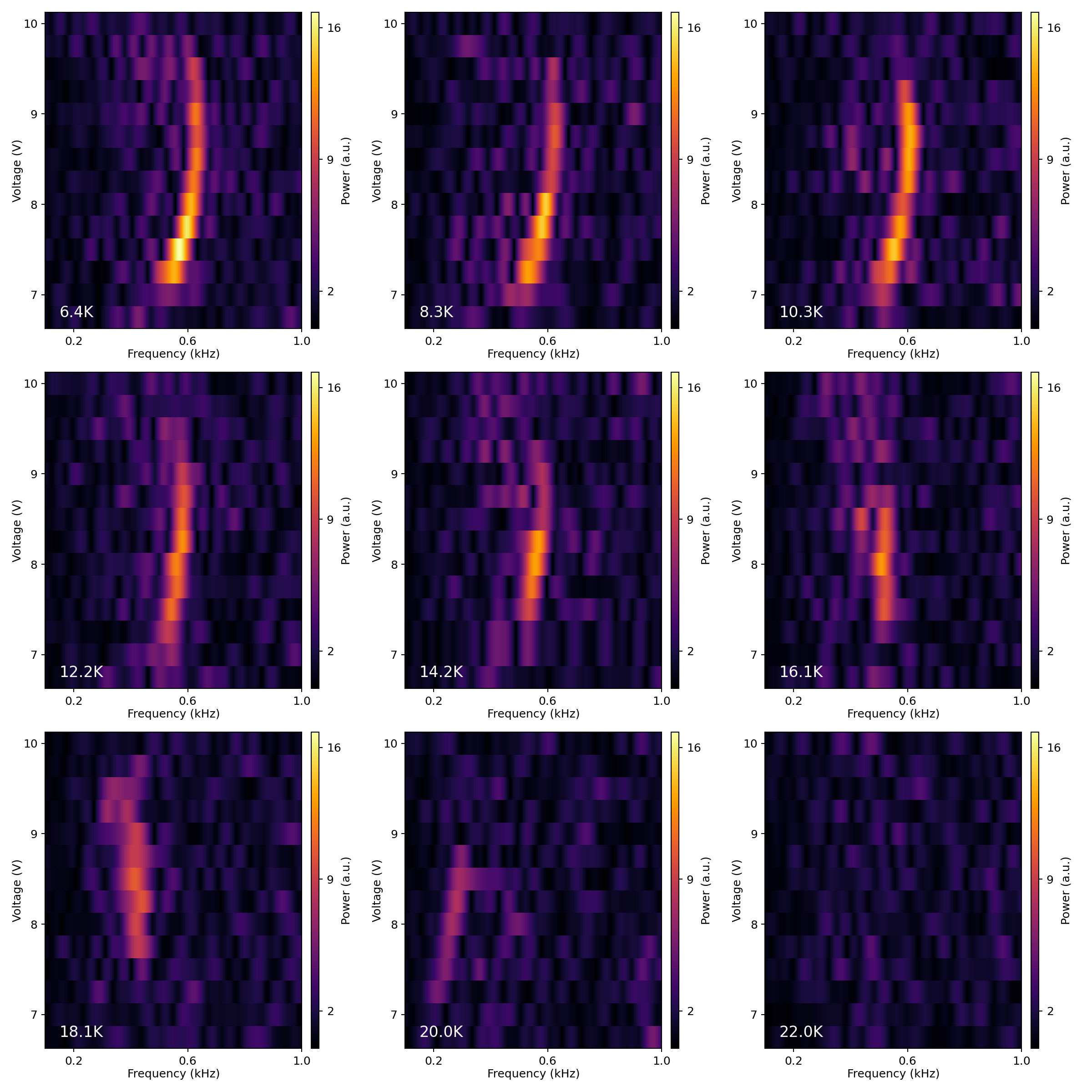} 
\caption{\textbf{FFT of T center EL oscillations per voltage and temperature.} Complete dataset from Main Text Fig. 4a.}
\label{SIFig1}
\end{figure*}

\newpage

\section{Complete Frequency Map of Diode Current Phase Diagram}
Here we present the extended data used to generate the phase map of Main Text Fig. 4b. The same process is repeated as described in SI Section I, but instead the time-trace data used to generate FFT maps is collected from the diode current measured on an oscilloscope, using the measurement circuit of Main Text Fig. 4f. The complete dataset is given in Supplementary Fig. \ref{SIFig2}.

\begin{figure*}[ht!]
\includegraphics[scale = 0.9]{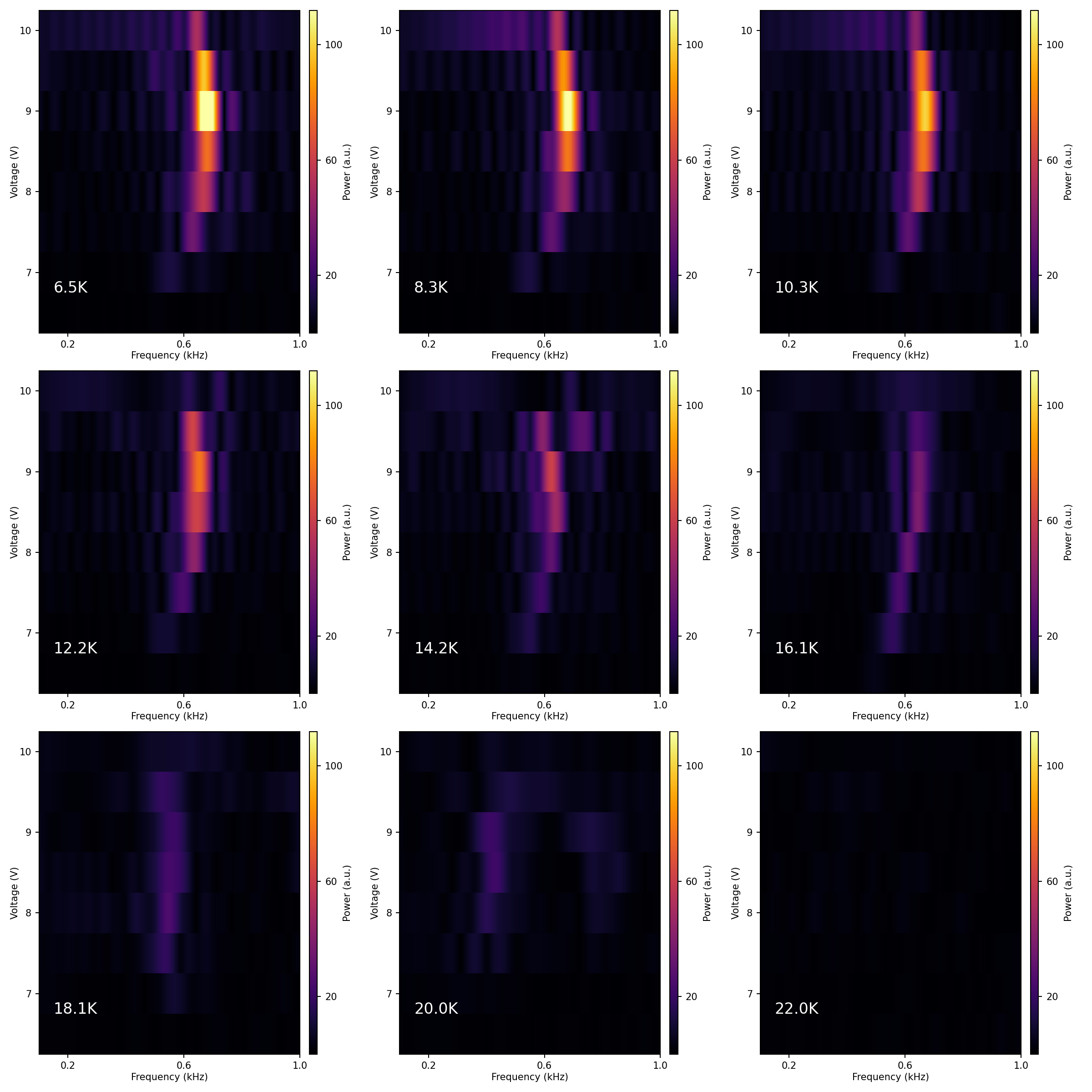} 
\caption{\textbf{FFT of diode current oscillations per voltage and temperature.} Complete dataset from Main Text Fig. 4b.}
\label{SIFig2}
\end{figure*}
\newpage

\section{Optical Schematic}
An optical schematic is given in Supplementary Fig. \ref{SIFig13} for the custom-built free-space scanning confocal microscope used to investigate the T center-integrated PIN-diode electroluminescence and photoluminescence at 6~Kelvin.  

\begin{figure*}[ht!]
\includegraphics[scale = 1]{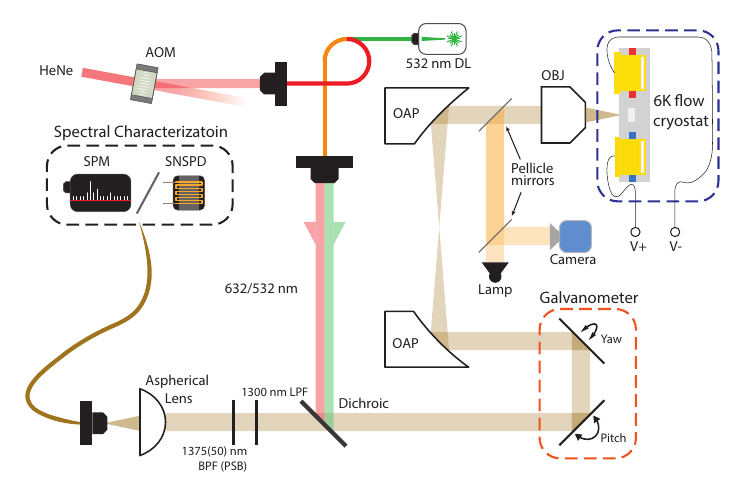} 
\caption{\textbf{Optical Diagram of Free-Space Confocal Microscope} A custom-built free-space confocal microscope is utilized to investigate the optical response of a PIN-diode integrated T center ensemble.}
\label{SIFig13}
\end{figure*}

The sample is mounted in a Janis ST-500 continuous-flow Helium cryostat. Bias is applied to the sample through the cryostat using a Keithley 2400 source meter for static bias, or a Rigol arbitrary waveform generator (AWG) for fast-switching bias (pulsed EL). Beam steering is performed with a galvanometer and a reflective 4f-relay using off-axis parabolic (OAP) mirrors--selected to operate across the broad range of wavelengths studied. Photons are collected through a series of long-pass and/or band-pass filters, depending on the experiment, and an O-band optimized aspherical lens. Emission is measured on a spectrometer, or superconducting nanowire single photon detectors (SNSPDs, PhotonSpot--housed on the 1~K stage of a dilution fridge), depending on the experiment. Optical pumping is incorporated via a long-pass dichroic mirror reflecting either a continuous-wave 532~nm excitation laser, or a pulsed 632~nm excitation laser (routed through an acousto-optic modulator for pulsed PL lifetime measurements). An imaging path is incorporated using a visible lamp, visible camera, and series of partially reflective pellicle mirrors. 

\clearpage

\section{Measurement Circuit Diagram}
\begin{figure*}[ht!]
\includegraphics[scale = 1.2]{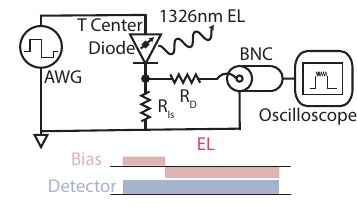} 
\caption{\textbf{Measurement circuit diagram} A transient positive-bias square waveform with otherwise constant reverse bias (bottom) is generated with an AWG and delivered to the PIN-diode housed in a continuous-flow cryostat. The embedded T center emission is collected with the optical schematic shown in SI Section III, where transient optical data is collected with SNSPDs, while the current is simultaneously readout on a current sensing resistor $R_IS$. $R_D$ enables integration of a BNC and oscilloscope while decoupling the load. }
\label{SIFig_circuit}
\end{figure*}

A circuit is constructed to simultaneously probe correlations in optical (T center) and electrical (diode) transient oscillations (Supplementary Fig. \ref{SIFig_circuit}). Diode current is measured on a low-ohmic current-sensing resistor $R_{IS}$ and read-out on an oscilloscope. Pulse sequence for transient oscillations analysis (bottom).

A $30\mu s$ positive bias is applied with a variable amplitude, followed by a continuous reverse bias of -8.5V (see SI Section XII for pulse-sequence optimization). Photons are collected on SNSPDs for the duration of the $150\mu s$ pulse period. 

\clearpage

\section{Investigation of Background Emission}
\begin{figure*}[ht!]
\includegraphics[scale = 1]{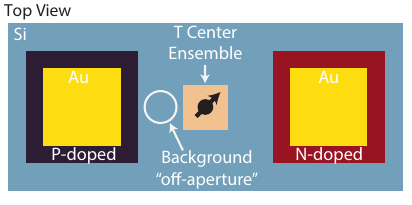} 
\caption{\textbf{Electrically active background defects.} The T center ensemble is localized to a 50~$\mu$m region of the PIN-diode at the center of the I-region. Outside of the implant aperture, there are no T centers present. Nonetheless, optically-active background defects fluoresce under both optical- and electrical-pumping. By moving the collection optics off-aperture, we can uniquely isolate the behavior of the sample distinct from the T center ensemble.}
\label{SIFig3}
\end{figure*}

Due to the lithographically-defined spatially-confined nature of the T center ensemble in our devices, it is possible to investigate the electrical response of the background--defined as the region spatially offset from the T center implant aperture (Supplementary Fig. \ref{SIFig3}). In this manner, we can isolate and separate the behavior we see while measuring \textit{on} the T center ensemble, to \textit{off} it.

In this background region, we find optically active defects which fluoresce under electrical pumping. In the following subsections we show that this background both (1) does not oscillate in response to the identical conditions explored in the main text which induce T center oscillations (Supplementary Fig. \ref{SIFig4}), and (2) that these defects posses a distinct excited state lifetime (Supplementary Fig. \ref{SIFig5}). 

\subsection{Absence of Background Oscillations}
\begin{figure*}[ht!]
\includegraphics[scale = 1.2]{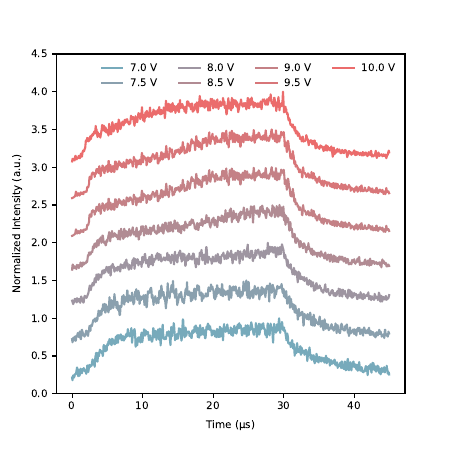} 
\caption{\textbf{Pulsed Electroluminescence of Background.} Using identical protocol and conditions which induce the T center EL and diode current oscillations of Main Text Fig. 4, we characterize the optical emission of background defects. We establish that the background \textit{does not oscillate}, but instead exhibits a time-trace of emission which roughly matches the applied square wave.}
\label{SIFig4}
\end{figure*}

To establish that the oscillations observed in the optical signal measured under pulsed electroluminescence of Main Text Fig. 4 in-fact exclusively emanate from the T center ensemble, we investigate the optical signal collected off-aperture (background) under identical pulse conditions. In addition, we search for oscillations under strict spectral filtering about the T center spectrum, described in SI Section VI. Notably, we find that the background emission does not oscillate under any field condition (Supplementary Fig. \ref{SIFig4}). 

Though this confirms that the T centers uniquely oscillate, this finding is in fact quite surprising. In Main Text Fig. 4b we already established that the \textit{current} in the diode is oscillating proportionally to the T center spectra. However this data reveals that the background emission intensity instead simply follows roughly the shape of the applied square wave, and does not exhibit the induced oscillatory component which the T centers do. This distinction could be explained by the extended lifetime of the background emitters.

\subsection{Electroluminescence Lifetime of Background}
In Supplementary Fig. \ref{SIFig5}, we characterize the excited state lifetime of these background defects under fast electrical switching and find a decay which is distinct from that of the T centers measured in Main Text Fig. 2. This decay time of $2.39~\mu s$ is $5\times$ longer than that measured on the T center ensemble using identical conditions ($1.5~\mu s$ square-wave 10~V pulse, with a $\sim100ns$ electrical turn-off switching time). 

\begin{figure*}[ht!]
\includegraphics[scale = 1.2]{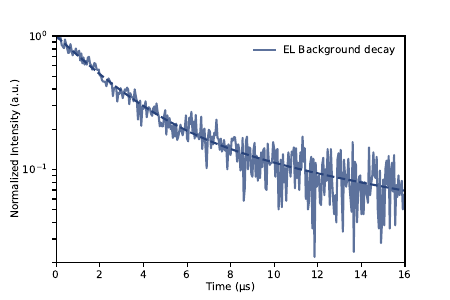} 
\caption{\textbf{Electroluminescence Lifetime of Background Emission.} An excited state lifetime of $2.39~\mu s$ is measured from the background defects under identical fast electrical switching utilized to characterize the T centers in Main Text Fig. 2. This slow decay is on the order of the EL slow-component observed in SI Fig. \ref{SIFigDecay}.}
\label{SIFig5}
\end{figure*}

\clearpage

\section{T Center Oscillations under Spectral Filtering of T Center Phonon Sideband}
\begin{figure*}[ht!]
\includegraphics[scale = 1]{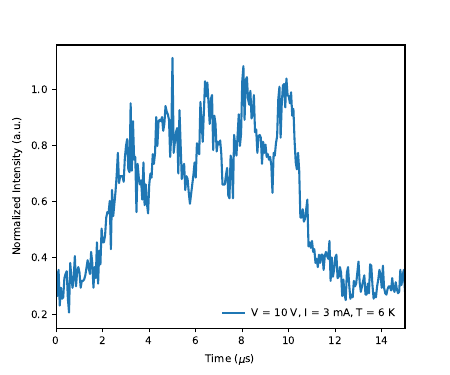} 
\caption{\textbf{EL Oscillations collected with 1375±25~nm band-pass filter.} Electroluminescence measured on implant aperture with a 1375±25~nm band-pass filter on collection optics, confirming that the optical origin of observed oscillations is T center emission. A 10$\mu s$ square-wave is applied to the diode to induce the observed oscillations.}
\label{SIFig6}
\end{figure*}

In addition to distinguishing the optical origin of the observed oscillation by moving on and off the T center ensemble aperture--as described above in SI Section V--we characterize the observed oscillations on-aperture under spectral filtering. We add to the collection optics a 50~nm band-pass filter centered at 1375~nm (1375±25~nm) and repeat the measurement of Main Text Fig. 4a. Notably, we find that the oscillations are still strongly observed, and in fact that the amplitude intensity of oscillation is modulated by approx. 50\% (Supplementary Fig. \ref{SIFig6}). 

In this way, we establish that the optical source of the observed oscillatory signal is indeed exclusively the T center ensemble. Interestingly, the spectrally-filtered oscillation amplitude peak-to-trough intensity varies by roughly $10\times$ that of the diode current--indicating an amplified sensitivity to the oscillatory electrical dynamics.

\clearpage

\section{Excited State Lifetime Decays}
The PL and EL T center excited state lifetime decays are investigated under pulsed optical and electrical excitation in Supplementary Fig. \ref{SIFigDecay}, where comparable decay times of 522±2~ns (PL) and 481±11~ns (EL) are observed in the fast decay components of a bi-exponential fit. While these excited state lifetimes are on the order of PL lifetimes reported in literature, the difference could be attributed to non-radiative pathways induced from insufficient lattice healing in our flash anneal. We also observe the presence of a slow decay component in the EL which could potentially be explained by the existence of electrically-active intrinsic background defects in the silicon, overlapping spatially and spectrally with the T center--as discussed above in SI Section V.

\begin{figure*}[ht!]
\includegraphics[scale = 1.5]{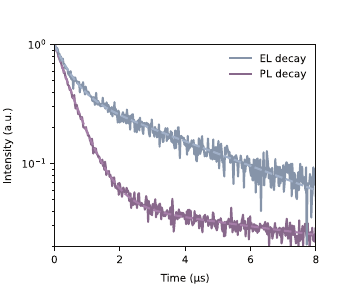} 
\caption{\textbf{Variation in PL and EL excited state decay} Excited state lifetime decays of T center under PL and EL excitation, revealing similar short decay components of 522(2) ns (PL) and 481(11) ns (EL), with distinct slow decay components, attributed to electrically-active background defects.}
\label{SIFigDecay}
\end{figure*}

\clearpage

\section{Electronics Switching Speed}
It is essential that the rate of decay of the diode is much faster than that of the emitter, in order to correctly observe the emitter EL lifetime. In Supplementary Fig. \ref{SIFig_switch}, we directly measure and compare the T center lifetime against the didoe current switching speed. This measurement confirms a fast turn-off time of approx. 100~ns, significantly faster than the emitter. 

\begin{figure*}[ht!]
\includegraphics[scale = 1.2]{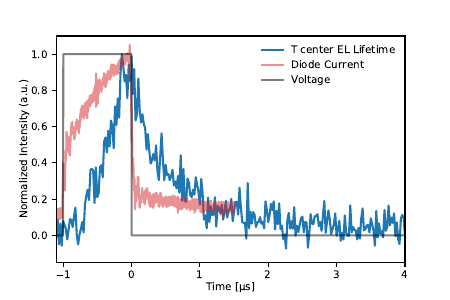} 
\caption{\textbf{Comparison of T center EL lifetime to diode turn-off time} T center electroluminescence is measured with a $1\mu$s $10~V$ square-wave applied to the diode. The diode current is directly measured with the circuit of Supplementary Fig. \ref{SIFig_circuit}. The diode turn-off time is compared against the EL decay time of the T center, revealing very fast switching (approx. 100~ns) which does not interfere with the emitter decay. }
\label{SIFig_switch}
\end{figure*}

\newpage

\section{Temperature-Dependent Oscillation Frequency Damping}
The maximum oscillation frequency observed at a particular field is plotted as a function of temperature in Supplementary Fig. \ref{SIFig_OscFreq}. This data is extracted from the phase map generated in Main Text Fig. 4c-d. The frequency becomes gradually reduced until the oscillation-supporting phase space is exited and the electric field transitions back to DC. 

\begin{figure*}[ht!]
\includegraphics[scale = 1.2]{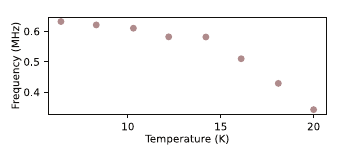} 
\caption{\textbf{Oscillation Frequency Decay} Maximum observed oscillation frequency as a function of temperature.}
\label{SIFig_OscFreq}
\end{figure*}

\clearpage

\section{Oscillation Quenching under pulsed forward bias and continuous above-band illumination}
Here we consider the behavior of the T center EL oscillations under CW 532~nm illumination. We postulate that by first finding a set of conditions which yield stable EL oscillations, the phase transition can be modified by directly altering the order parameter via optical injection of electrons and holes under above-bandgap optical pumping. In Supplementary Fig. \ref{SIFig7}, we sweep the applied optical power to the junction conditions which support pulsed EL T center oscillations. We indeed find that at high optical power the oscillations become suppressed, instead favoring constant bright T center emission. Interestingly however, at low applied optical powers ($10-20~\mu~W$), the oscillations appear stronger than at no optical power. This suggests that the combination of electron concentration and/or impact ionization coefficient is briefly modified to \textit{improve} the stability of the phase transition under optical injection of carriers, before ultimately suppressing it. 

\begin{figure*}[ht!]
\includegraphics[scale = 1]{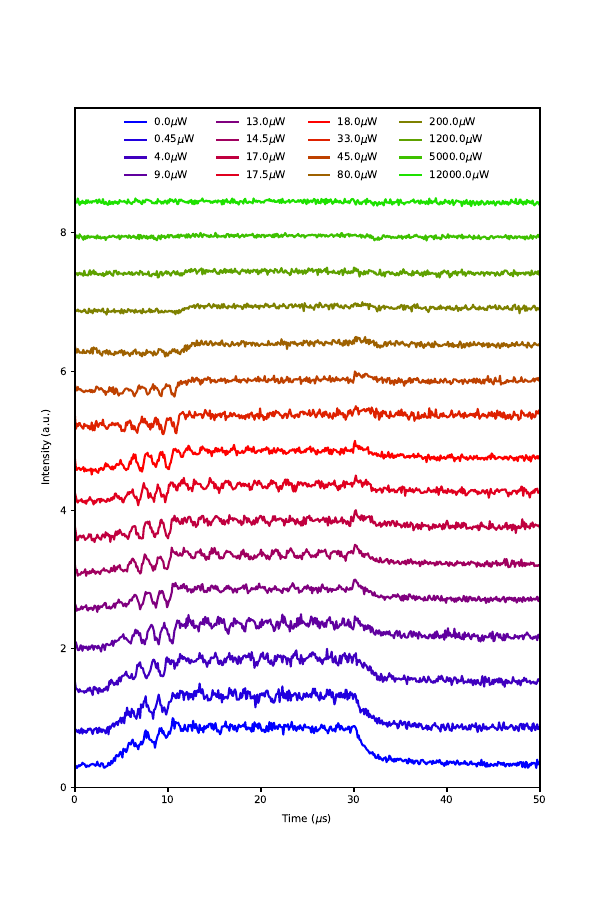} 
\caption{\textbf{EL oscillation quenching under CW illumination.} EL oscillations are suppressed at increased power of CW above-bandgap illumination.}
\label{SIFig7}
\end{figure*}

\newpage

\section{T center and Free Exciton Emission Saturation}
In Supplementary Fig. \ref{SIFig8}, we characterize the saturation curves of the T center ensemble under optical and electrical injection (left), and the emission intensity of the silicon free exciton (FE) on- and off- aperture (right). 

\begin{figure*}[ht!]
\includegraphics[scale = 1]{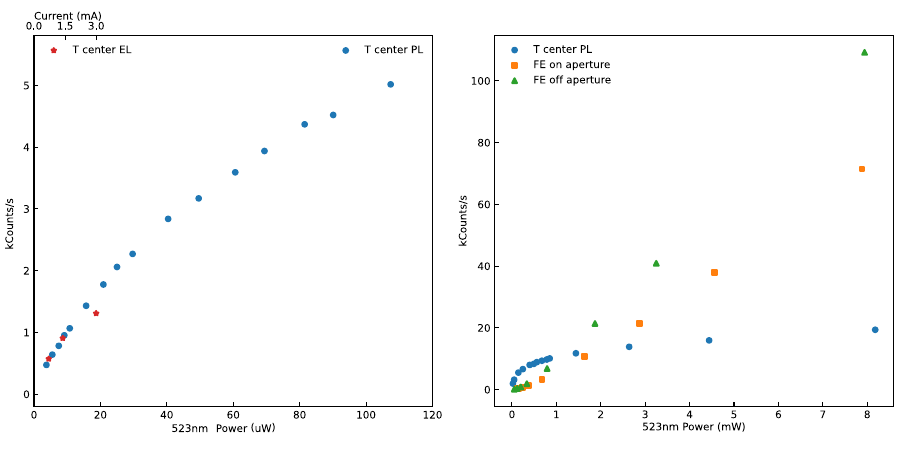} 
\caption{\textbf{Silicon Free-Exciton (FE) and T center ensemble optical emission intensity at 6~Kelvin.} (Left) comparison of T center PL and EL emission intensity. A 532~nm laser is utilized for PL. (Right) T center saturation and FE emission intensity as a function of input optical power.}
\label{SIFig8}
\end{figure*}

We first investigate the optical pumping of the T center ensemble at low input power, and compare the intensity to that under electrical injection (Supplementary Fig. \ref{SIFig8} Left). We observe that the EL intensity does not saturate in the regime we investigate (up to 3mA, 10V), and that the intensity closely follows that of optical-injection curve up to approx. $25uW$. We do not investigate higher input currents to ensure safe operation of the device, though complete characterization of EL saturation could be explored under pulsed high-current regimes to prevent heating. 

In Supplementary Fig. \ref{SIFig8} Right, we now compare the high input optical power pumping of the T center ensemble to the silicon FE on and off-aperture. We do observe saturation of the T center ensemble under high-power optical pumping, and interestingly observe that the FE is suppressed on-aperture relative to off-aperture. Further, at low input powers ($\leq 1mW$), T center emission dominates that of FE emission. Considering the recombination mechanism of T center excitation being the capture of FE which become bound to the defect site and contribute to T center luminescence, it is interesting to consider whether this could be understood as participatory FEs are contributing to T center emission rather than freely recombining and fluorescing. 

\newpage

\section{Impact of Reverse Bias on T center Oscillations}
During optimization of the applied pulse sequence to induce EL oscillations, we began by exploring switching the square-wave from 0~V to the high-point. We observed a gradual decrease in the oscillation amplitude over repeated cycles, and postulated that this could be due to the accumulation of charge in the diode and/or contacts. To reset the charge in the junction per switching cycle, we evaluated applying a reverse-bias during the ``off" period of the pulse-sequence, and found that the oscillation strength was recovered. A brief summary of these findings is presented in Supplementary Fig. \ref{SIFig9}, where for the Main Text we implemented these findings and exclusively utilized a pulse sequence which holds a constant -8.5~V reverse-bias during the ``off" time of the switched-EL pulse sequence.  

\begin{figure*}[ht!]
\includegraphics[scale = 1]{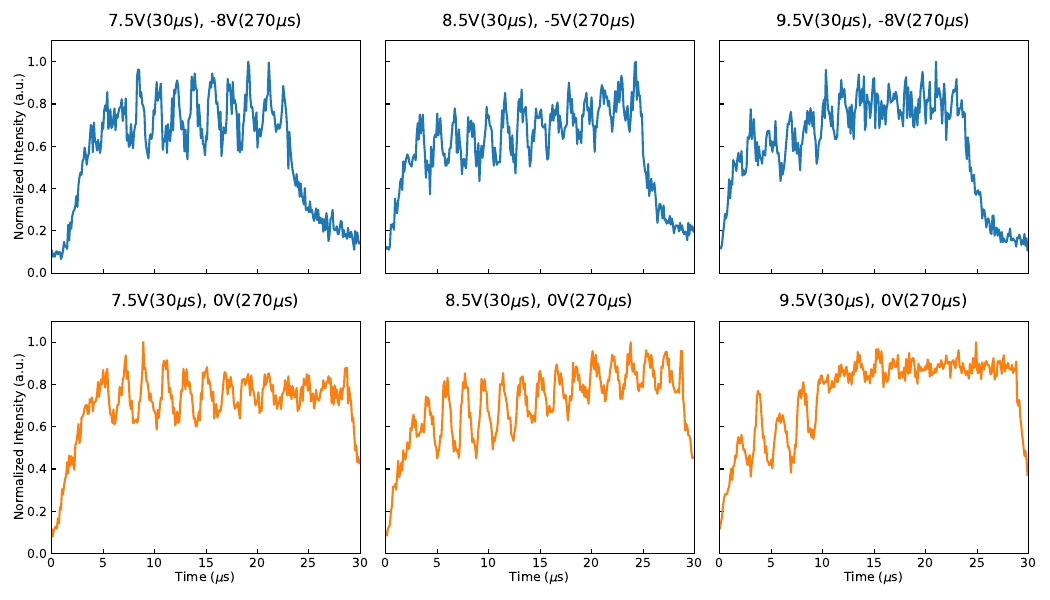} 
\caption{\textbf{Oscillation Impact of Reverse Bias in EL Pulse Sequence.} Comparison of varied pulse conditions in optimizing induced EL oscillations from a a $30~\mu s$ square wave, with reverse-bias held during the $270~\mu s$ ``off"-time of the sequence period (top), versus 0~V (bottom).}
\label{SIFig9}
\end{figure*}
\newpage
\newpage

\section{IV-curves under varied optical pumping}

In Supplementary Figs. \ref{SIFig10}-\ref{SIFig11}, we explore the impact of above-bandgap optical pumping on carrier concentrations, evaluated via IV curves at 6~Kelvin.  

First, in Supplementary Fig. \ref{SIFig10}, we find that optical illumination of the PIN-diode \textit{suppresses} the electron concentration above a voltage threshold (relative to no laser on), and that the IV-curves under both 532~nm and 632~nm laser illumination behave similarly. 

\begin{figure*}[ht!]
\includegraphics[scale = 1]{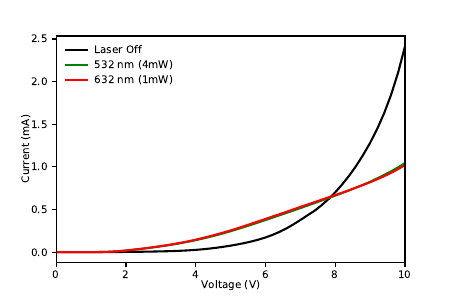} 
\caption{\textbf{Role of varied optical pumping on 6~Kelvin IV Curves.} Both red and green optical pumping suppress electron concentration in the diode above 8V.}
\label{SIFig10}
\end{figure*}

Next in Supplementary Fig. \ref{SIFig11}, we further explore the impact of CW laser illumination by considering the spatial position of optical pumping into the diode--where Center, Left, Right, Top, and Bottom refer to the position of the illumination laser spot relative to the T center ensemble aperture. 

\begin{figure*}[ht!]
\includegraphics[scale = 1]{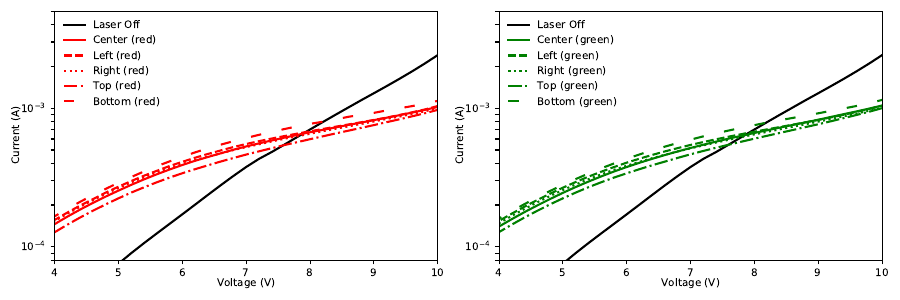} 
\caption{\textbf{Detailed analysis of optical pumping in 6~Kelvin IV curves.} Evaluation of the wavelength- and spatial dependence on optically-induced current suppression in the diode. All positions and both laser colors consistently yield a suppression in diode current above 8~V. Laser illumination positions referenced relative to T center aperture.}
\label{SIFig11}
\end{figure*}

We find that the position of optical injection makes no significant impact on the current dynamics--so long as optical pumping occurs within the diode ``I-region", and thus depletion region. However, we observe that above a certain voltage threshold--and thereby local field--the current is consistently suppressed relative to the case of no optical pumping. Though this finding warrants further study and cannot yet be fully understood, we note that the region at which the electron concentration drastically changes correlates with the voltage region of stable oscillations reported in the main text--and thus could be related to the low-temperature negative differential resistance. 

\newpage

\section{Cryogenic Electronics}
\begin{figure*}[ht!]
\includegraphics[scale = 1]{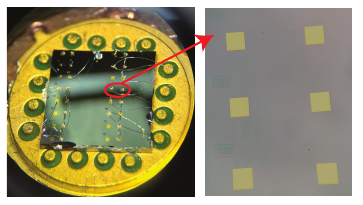} 
\caption{\textbf{Cryo-packaged Wirebonded Sample.} Close-up image of cryostat mounting of T center-integrated PIN-diodes. The sample and cryostat are integrated into a free-space scanning confocal microscope for optical characterization, compatible with simultaneous cryogenic electrical measurements.}
\label{SIFig12}
\end{figure*}

As described in Main Text Methods, during device fabrication and emitter synthesis, a 6-inch wafer is diced into $8\times8mm$ samples for cryogenic measurement. These samples are mounted into a 16-pin electrically-wired insert situated in the copper cold-finger of a Helium flow-cryostat (Supplementary Fig. \ref{SIFig12}). The fabricated diodes are wire-bonded to the cryostat insert, and electrical connection is externally applied. Each sample fits 16 separate PIN-diodes, with a T center ensemble integrated into the center of each. The cryostat is cooled using liquid Helium, and simultaneous optical and electrical characterization is enabled using a free-space scanning confocal microscope. 

\newpage

\section{DC-Bias T Center Frequency Shift}
The PL optical response of the T centers is characterized by reverse biasing the diode from 0 to -200~V (with diode breakdown occurring at -180~V under above-bandgap illumination), in Supplementary Fig. \ref{SIFigDCPL}. Below diode breakdown, up to a field strength of approximately 400~kV/m, no significant changes in emitter ensemble brightness, center wavelength, or linewidth are found--a shift of roughly 300~MHz at 140~kV/m is observed with no change in emission intensity. Above diode breakdown, leakage current induces milliwatts of power dissipation across the junction thereby heating the ensemble and quenching the optical emission. 

\begin{figure*}[ht!]
\includegraphics[scale = 1]{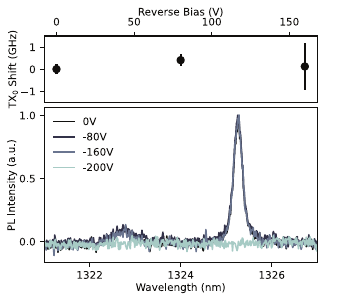} 
\caption{\textbf{T center Ensemble TX$_0$ PL Frequency Response} PL-pumped optical response of T center under reverse bias from 0 to -200V. The ensemble TX0 and TX1 lines show negligible changes in intensity, linewidth, or center frequency below the diode breakdown (up to -180V), where current-induced heating extinguish the lines above breakdown.}
\label{SIFigDCPL}
\end{figure*}

\newpage

\section{Band Diagram}
A conceptual band diagram under forward bias is reported in Supplementary Fig. \ref{SIFigBands}. The T center defect level is predicted to lie near the conduction band edge. Electroluminescence is observed under forward-bias, mediated by capture of free-excitons generated via (1) electron-hole current injection and/or (2) impurity-mediated impact ionization.

\begin{figure*}[ht!]
\includegraphics[scale = 1]{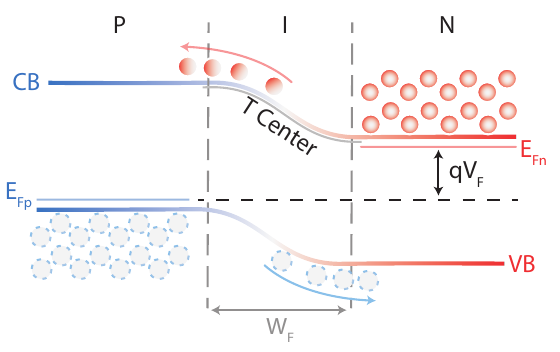} 
\caption{\textbf{PIN-Diode Biased Band Diagram} Electron and hole current is efficiently passed across the PIN junction when the band barrier $qV_f$ is lowered via application of forward bias. The T center trap level is predicted to lie near the conduction band edge, and the ensemble is localied to the I-doped region of the device.}
\label{SIFigBands}
\end{figure*}

\newpage

\subsection{Proper temperature stabilization in a Janis ST-500}
For data taking in Main Text Fig. 4, it was imperative to properly thermalize the emitter-diode system at a stable temperature for extended periods of time. During investigation we found that simply heating the cryostat using a temperature controller did not ensure thermal equilibrium of the T center ensemble and PIN-diodes, as the nature of a continuous-flow Helium cryostat can render efficient cooling in sub-regions of the system. We observed this via oscillations appearing stronger than they should at a given temperature due to more efficient cooling than the heater could combat. This was mitigated by balancing the rate of helium flow to the rate of applied heat from the cryo-heater--where we feedback on the necessary heat-load applied to the system and the measured cryostat temperature. 

Put simply, rapidly flowing helium will cool the system more efficiently than applying $100\%$ power from the cryo-heater can heat it, and thus results in pockets of regions hotter and colder than intended. To mitigate, we reduce He flow rate and reduce applied power until each are stable at a desired temperature. \\

Our protocol for temperature stabilization is as follows: 
\begin{enumerate}
    \item Enable PID-controlled heater in LakeShore 321 Temperature controller. This applies heat directly to the base of the cryostat cold-finger on which the diode sample is mounted. 
    \item Observe power-percentage applied from the heater. Do not let this value exceed roughly $10\%$ to reach desired cryostat base temperature.
    \item If heater power exceeds roughly $10\%$, reduce rate of He flow to the cryostat. 
    \item Monitor change in applied heater power and corresponding response in cryostat base temperature. 
    \item Wait approx. 10min to ensure system is stable (temperature and applied power do not fluctuate) and to let system thermalize to new temperature set-point. 
    \item Proceed with conducting experiment at new elevated temperature. 
\end{enumerate}
\newpage

\section{Extended discussion}
\subsection{Carrier Dynamics}
\textit{Deep level defects}, with energy levels far removed from the conduction or valence band edges can serve as traps having substantial trapping lifetimes. Shockley-Read-Hall (SRH) recombination is a well characterized means of non-radiative recombination \cite{shockley1952,hall1952}. It is not surprising, then that a number of early studies on $p^+ - i - n^+$ diodes in Au-doped Si identified NDR in the current-voltage characteristics \cite{lampert1961,weber1970, deuling1970,kassing1975,dudeck1977,jager1986,baumann1986,weman1989,symanczyk1991}. These studies focused on Au-doped high purity Silicon, where the amphoteric Au deep level can serve as both acceptor, in its negatively charged state, and as donor in its positively charged state \cite{weber1970}. As was noted by \cite{weber1970}, non-monotonic current-voltage behavior could be explained by the fact that ``the current is limited by recombination kinetics rather than mobilities" where the interplay of electron and hole recombination with the deep-level defect produced the NDR-like behavior. 

There are some important additional considerations for our measurements: (1) At the low temperatures for which our data are taken ($<$ 10K), the hole current density is significantly lesser than that of electrons, due to differing carrier concentrations and mobilities \cite{ottaviani1975,canali1975}. (2) The low rate of electron injection into the T-center region should provide high sensitivity in mapping out details of carrier capture and trap lifetimes. (3) Photo-excitation, in addition to electrical injection, dramatically changes the concentrations of electrons and holes. In addition, photo-excitation will produce equal concentrations of electrons and holes, yet the overall current density, $\vec{J}$ will still consist largely of electrons. It is therefore interesting to trace the forward-biased NDR dynamics as the PL intensity varies. Indeed, we investigate this condition in Supplementary Section X and find a suppression of the oscillations at increased PL intensity, whereby the order parameter--electron concentration--is modified by optical injection. 
For simplicity, we will treat the base temperature as $T=10K$. 

\subsection{Carrier Concentration}

The distinction between EL and PL intensity, and similarly the relative inefficiency of EL, can be understood by considering the carrier dynamics. High doses of implanted $^{11}$B and $^{31}P$  formed the P-type and N-type regions of the P-I-N diode. We can estimate the approximate temperature-dependent ionization of those dopant levels by assuming $E_a$, the energy level of the Boron acceptor is 45 meV from the Si valence band edge and $E_d$, the energy level of the Phosphorus donor is 44 meV from the Si conduction band edge \cite{sze1968resistivity}. Then, for example the concentration of ionized donors, $N_d^+$, giving rise to electrons in the conduction band is: 
\begin{equation}
    N_d^+ = \frac{N_d}{1+e^{E_d/(k_B T)}}
\end{equation}
where $k_B$ is Boltzmann`s constant. Thus, at low temperatures, there is a very low concentration of electrons and holes in the conduction and valence bands, respectively. Moreover, the concentration of holes is about an order of magnitude lower than that of electrons.

At 10~K there is a very sparse current density of approximately $1.6\times10^2$  electrons/cm$^{2}$/sec, assuming an electric field of ~ $1.2\times10^2$ V/cm (see below). Furthermore, the electron and hole mobility should differ at 6~K \cite{ottaviani1975,canali1975}. Given these differences in electron and hole density, it is necessary to consider how these dynamics could explain our observed electrically-injected T center emission. It is suggested that T center PL is mediated by capture of free excitons at the defect site \cite{bergeron2020}. The electrically-mediated formation of excitons necessary for T center emission could thus be formed via impact ionization \cite{eitan1982,maes1990}, whereby the increased relative electron density could collide with local impurities to ionize electron-hole pairs resulting in T center capture and excitation. Indeed, carbon is uniformly densely implanted in our samples and could serve as such a local impurity \cite{sugiyama2012}. Furthermore, the presence of a slow-decay component in the EL lifetime, yet absent from the PL lifetime, could be due to this distinct excitation mechanism of impact ionization, whereby an energetic charge carrier excites a charge neutral defect thus forming a bound exciton with relatively long lifetime--carbon related neutral defect complexes have indeed been investigated as hole traps in silicon \cite{kamiura1995,sugiyama2012}. Although our finding does not exclude the possibility of other emission mechanism, the correlation between the onset of EL and impact-ionization-mediated phase transition provides a solid evidence that the T center emit via a free exciton (FE) capture process. 

\subsection{Base temperature estimate of electron and hole density during oscillation}
We begin by estimating the electric field in the diode under the conditions which induce carrier oscillations. \\

For 7 Volts applied across an ``I-region" region of width $W=565~\mu m$, the electric field strength is: 

\begin{equation}
\vec{E}= 1.2\times10^{-2} V/cm
\end{equation}

The electron current density is then given by: 

\begin{equation}
\vec{J}=q\mu_n {\bf n}\vec{E}
\end{equation}

Letting $\mu_n=1350cm^{2}Vsec$, ${\bf n}=10^{-3}$ electrons/cm$^{3}$, and $\vec{E}=1.2\times10^{-2} V/cm$, we obtain: \\

\begin{center}
    $\vec{J}=1.6\times10^2$  electrons/cm$^{2}$/sec
\end{center}

Where values of $\mu_n$ and ${\bf n}$ at 10~K are approximated from literature \cite{streetmanbanerjee}, and given by our boron and phosphorus dopants at a concentration of $1\times10^{19}$ 

\bibliography{Ref.bib}